\documentclass{article}[12pt]
\usepackage{epsfig} 
\usepackage{cite} 
\newcommand{\labbel}[1]{\label{eq:#1}} 
\def\ie{{\it i.e.}\phantom{.}} 
\def\Im{\hbox{Im}} 
\def\F{F^{(2)}} 
\def\e{{\rm e}} 

\begin{document}

\unitlength1cm
\begin{titlepage}
\vspace*{-1cm}
\begin{flushright}
TTP03-05\\
%%hep-ph/0302162\\
%%February 2003
\end{flushright}
\vskip 3.5cm
\renewcommand{\topfraction}{0.9}
\renewcommand{\textfraction}{0.0}

\begin{center}
\boldmath
{\Large\bf Two-Loop Form Factors in QED
}\unboldmath
\vskip 1.cm
{\large P. Mastrolia}$^{a,b}$ and
{\large E. Remiddi}$^{a,c}$
\vskip 1.0cm

{\it $^a$ Dipartimento di Fisica,
    Universit\`{a} di Bologna, I-40126 Bologna, Italy \\ 
       \vskip 0.4cm
       $^b$ Institut f\"ur Theoretische Teilchenphysik,
            Universit\"at Karlsruhe, D-76128 Karlsruhe, Germany \\ 
       \vskip 0.4cm
       $^c$ INFN, Sezione di Bologna, I-40126 Bologna, Italy \\
}
\end{center} 

\vskip 2.5cm

\begin{abstract} 
We evaluate the on shell form factors of the 
electron for arbitrary momentum transfer and finite electron mass, 
at two loops in QED, 
by integrating the corresponding dispersion 
relations, which involve the imaginary parts known since a long time. 
The infrared divergences are parameterized in terms of a fictitious 
small photon mass. The result is expressed in terms of Harmonic 
Polylogarithms of maximum weight 4. The expansions for small and 
large momentum transfer are also given. 

\vspace*{1cm}
\noindent
{\it Key words:} Feynman diagrams, Multi-loop calculation, Dispersion
Relations \\
{\it PACS:} 11.55.Fv, 12.20.Ds, 13.40.Gp
\end{abstract} 
\vfill

\end{titlepage} 

\renewcommand{\theequation}{\mbox{\arabic{section}.\arabic{equation}}} 
%%%%%%%%%%%%%%%%%%%%%%%%%%%%%%%%%%%%%%%%%%%%%%%%%%%%%%%%%%%%%%%%%%%%%%%%%%%%
\section{Introduction} 
\label{sec:int} 
\setcounter{equation}{0} 

The imaginary parts of the on shell form factors of the electron 
at two loops in QED (the relevant graphs are shown for completeness 
in Fig.1) 
were calculated analytically long ago~\cite{BMR} in terms of 
Nielsen's polylogarithms~\cite{nielsen} of maximum weight $w=3$. 
The real parts however are still missing, as the integration of 
the dispersion relations for the form factors could not be carried 
out in closed form within the family of Nielsen's polylogarithms 
only. \par 
We show in this paper that the previous calculation can be completed 
within the family of the Harmonic Polylogarithms (HPL's) introduced 
in~\cite{hpl}. The result involves HPL's of maximum weight $w=4$. 
\par 
The plan of the paper is as follows. In Section 2 we recall, 
from~\cite{BMR}, the proper dispersion relations satisfied by the 
two form factors. In Section 3 we list the results of 
the analytic integration for arbitrary momentum transfer $t=-Q^2$ and
finite electron mass $m$. 
For positive $Q^2$ all the terms are real; as the expression is 
exact, it can be continued to $t$ timelike and above 
threshold, where the form factors develop their imaginary parts. 
In Section 4 we give 
the expansions of the form factors for $Q^2\to\infty$, and in 
Section 5 the expansions for $Q^2\to 0$. In the Appendix A
we recall the definition 
and the main properties of the HPL's, while in Appendix B we derive the 
analytic formulas which were used in the calculation. 

%%%%%%%%%%%%%%%%%%%%%%%%%%%%%%%%%%%%%%%%%%%%%%%%%%%%%%%%%%%%%%%%%%
\begin{figure}[h]
\includegraphics*[3.0cm,11.5cm][25cm,20.5cm]{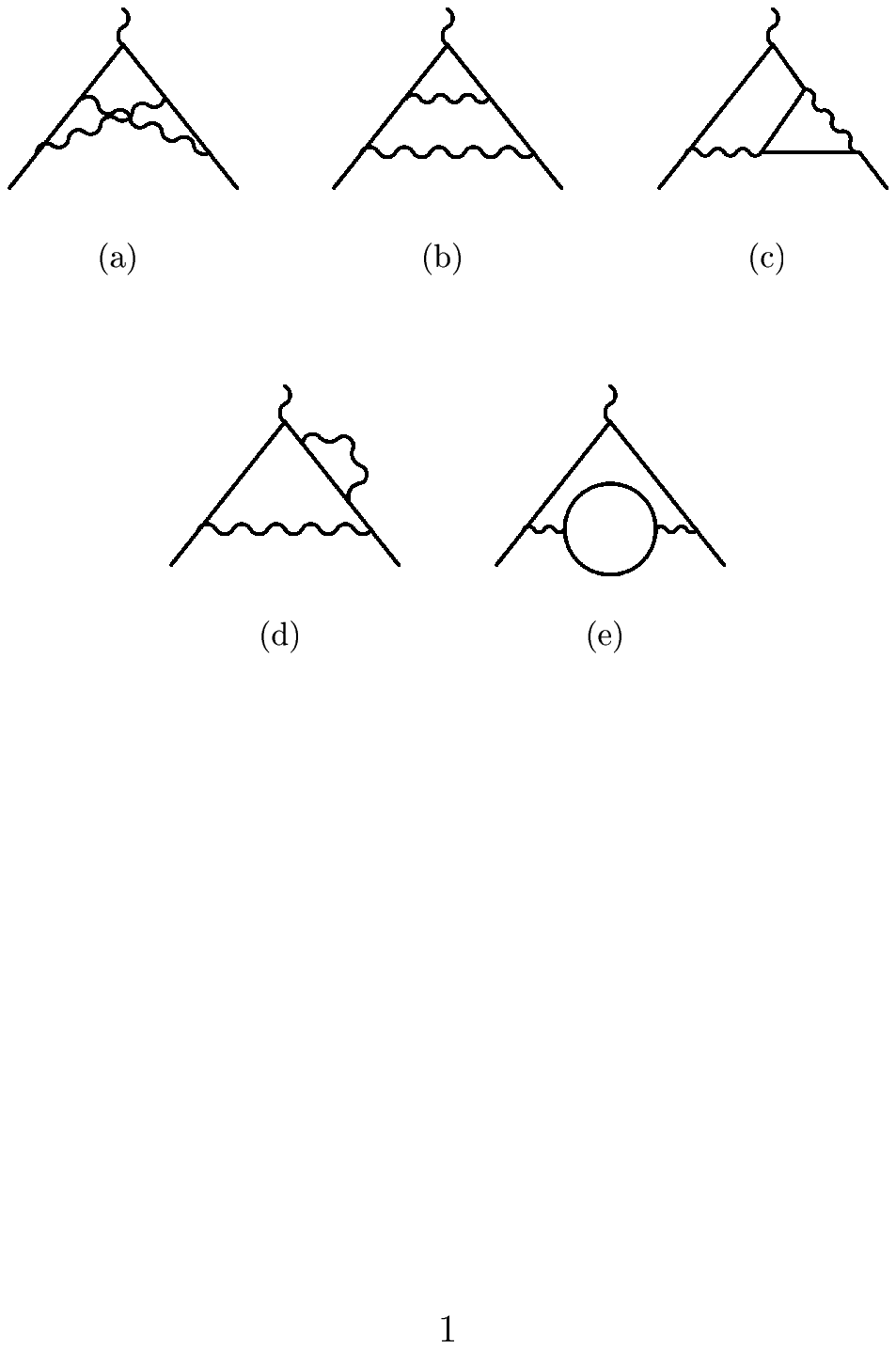}
\caption{The Vertex Graphs at two loops in QED (multiplicities understood). }
\end{figure}
%%%%%%%%%%%%%%%%%%%%%%%%%%%%%%%%%%%%%%%%%%%%%%%%%%%%%%%%%%%%%%%%%%

%%%%%%%%%%%%%%%%%%%%%%%%%%%%%%%%%%%%%%%%%%%%%%%%%%%%%%%%%%%%%%%%%%%%%%%%%%%%
\section{Dispersion Relations}
The imaginary parts of the two 
on-shell vertex form factors of the electron at two loop in QED, 
$\Im F_1^{(2)}(t)$, $\Im F_2^{(2)}(t)$ were evaluated in~\cite{BMR} 
for arbitrary value of the momentum transfer $t$ and finite electron
mass, 
within the Pauli-Villars regularization scheme for the ultraviolet 
divergences (needed for the renormalization of the inserted one 
loop subgraphs) and by using a small fictitious mass $\lambda$ for 
the regularization of the infrared divergences. 
In the imaginary parts $t$ is above the threshold $4m^2$, where $m$ 
is the electron mass; by introducing the dimensionless variable $x$ 
through 
\begin{equation} 
     t = m^2\frac{(1+x)^2}{x}\ , \quad \quad \quad 
     x = \frac{\sqrt{t}-\sqrt{t-4m^2}}{\sqrt{t}+\sqrt{t-4m^2}} \ , 
\labbel{defx} 
\end{equation} 
all the imaginary parts are can be written in terms of $x$, $\log{x}$, 
$\log(1-x)$, $\log(1+x)$ and Nielsen polylogarithms~\cite{nielsen} 
of maximum weight 3 and of argument $x, -x, x^2$. 
\par 
As discussed in Eq.(1.30) of~\cite{BMR}, the corresponding real parts are 
then given by properly subtracted dispersion relations, which for 
spacelike momentum transfer $t=-Q^2$ read 
\begin{eqnarray}
 F_1^{(2)}(-Q^2) &=& 
     - \frac{4 m^2 Q^2}{Q^2 + 4m^2} \ {F_1^{(2)}}'(0) \nonumber\\ 
     &-& \frac{Q^4}{Q^2+4m^2} \int_{4m^2}^\infty
                \frac{dt'}{t' + Q^2} \ \frac{t' - 4m^2}{t'^2}
                  \ \frac{1}{\pi}{\mbox{Im}} \ F_1^{(2)}(t') \ ,
\labbel{ReF1}
\end{eqnarray}
\begin{eqnarray}
F_2^{(2)}(-Q^2) &=&  \frac{4 m^2}{Q^2+4m^2} \ F_2^{(2)}(0) \nonumber\\ 
     &+& \frac{Q^2}{Q^2 + 4m^2} \int_{4m^2}^\infty
                \frac{dt'}{t' + Q^2} \ \frac{t' - 4m^2}{t'}
                  \ \frac{1}{\pi} {\mbox{Im}} \ F_2^{(2)}(t') \ . 
\labbel{ReF2}
\end{eqnarray}    
Indeed, the above imaginary parts are singular at 
$t=4m^2$, due to the infrared singularities of the 2-photon ladder graph 
showing up in the $\lambda\to0$ limit -- hence the subtraction at 
$Q^2=-4m^2$ and the factor $(t'-4m^2)$ for making the dispersive integrals 
convergent at threshold. Note that the charge slope ${F_1^{(2)}}'(0)$ 
and the magnetic anomaly $F_2^{(2)}(0)$ are needed for writing the 
above dispersion relations (which therefore cannot be used, in that 
form, for their calculation). Let us further observe that 
Eq.(\ref{eq:ReF1}) gives $F_1^{(2)}(0)=0$, the usual renormalization 
condition of the charge form factor. 
\par 
Similarly to Eq.(\ref{eq:defx}), in the spacelike case we put 
\begin{equation} 
     Q^2 = m^2\frac{(1-y)^2}{y}\ , \quad \quad \quad 
     y = \frac{\sqrt{Q^2+4m^2}-\sqrt{Q^2}}{\sqrt{Q^2+4m^2}+\sqrt{Q^2}} \ ; 
\labbel{defy} 
\end{equation} 
with the change of integration variable $t' = m^2(1+x')^2/x'$, the 
dispersive integrals occurring in Eq.s(\ref{eq:ReF1},\ref{eq:ReF2}) are 
seen to be of the form 
\begin{equation} 
  \int_{4m^2}^{\infty} \frac{dt'}{t'+Q^2} f(t') 
 = \int_0^1 dx' \left( \frac{1}{x'} - \frac{1}{x'+y} - \frac{1}{x'+1/y} 
                \right) \phi(x') \ . 
\labbel{dxy} 
\end{equation} 
When the actual explicit expressions of the imaginary parts in terms of 
Nielsen's polylogarithms are used within the dispersive integrals 
Eq.(\ref{eq:ReF1},\ref{eq:ReF2}), it is found that the resulting 
integrals cannot be evaluated in terms of that same class of functions. 
The Nielsen's polylogarithms, however, are special cases of the 
Harmonic Polylogarithms~\cite{hpl}, HPL's; when the more general formalism 
of the HPL's is used, one finds that not only the imaginary parts, 
but also the real parts given by the dispersive integrals can all be 
expressed in terms of HPL's of argument $y$ and 1 (the HPL's of 
argument 1 are expressed in turn as combinations of a few mathematical 
constants like $\zeta(3)$, the Riemann $\zeta$-function of index 3, 
$\pi^2$ and $\ln2$). 

We give in the following paragraphs the explicit results of the 
integration. The relevant formulae are shortly discussed in the 
Appendix. 
%%%%%%%%%%%%%%%%%%%%%%%%%%%%%%%%%%%%%%%%%%%%%%%%%%%%%%%%%%%%%%%%%%%%%%%%%%%%

\section{Results}

We present in this section the results of the analytic integration 
of Eq.(\ref{eq:ReF1}) and Eq.(\ref{eq:ReF2}) as a function of 
$Q^2 = - t$. As already said, 
$\lambda$ is the ``small photon mass" used for the parameterization 
of the infrared divergences. 
All the polylogarithms depend on $Q^2$ through the 
variable $y$ defined in Eq.(\ref{eq:defy}). Let us recall that 
$y(Q^2=\infty)=0$, $y(Q^2=0)=1$; when the momentum transfer $t$ is 
positive (timelike) and varies from $0$ to the 
physical threshold $t=4m^2$, $Q^2=-(t+i\epsilon)$ varies from 
$-i\epsilon$ to $Q^2=-4m^2-i\epsilon$; correspondingly, 
$y=\e^{i\phi}$, with $\phi=0$ at $t=0$ and $\phi=\pi$ at $t=4m^2$ 
(and the form factors are still real despite the complex value 
of $y$). When $t$ is above the threshold, it is convenient to write 
$y=-x+i\epsilon$, where $x$ is the variable defined in 
Eq.(\ref{eq:defx}), so that as $t$ varies from $4m^2$ to $+\infty$, 
$x$ varies correspondingly from $1$ to $0$. For $y=-x+i\epsilon$ 
the polylogarithms with rightmost index equal to $0$ develop an 
imaginary part which generates the imaginary parts of the 
form factors (see ~\cite{hpl,num1hpl} and the Appendix A for more details). 
\par 
The explicit results are 
%%%%%%%%%%%%%%%%%%%%%%%%%%%%%%%%%%%%%%%%%%%%%%%%%%%%%%%%%%%%%%%%%%%%%%%%%%% 
\begin{eqnarray}
 \F_1(-Q^2) & = & 
         \ln^2 \Big( \frac{\lambda}{m} \Big) 
          \bigg\{ 
%%%%%%%%%%
         \frac{m^2}{Q^2} \bigg[ H(0,0;y) \bigg]
       + \frac{Q^2}{\sqrt{Q^2 (Q^2+4m^2)}}  \bigg[ H(0;y) \bigg]
       + \frac{m^2}{Q^2+4m^2}  \bigg[ - H(0,0;y) \bigg]
         \nonumber \\
  &  & 
%%%%
       \qquad \qquad \quad
       + \frac{m^2}{\sqrt{Q^2 (Q^2+4m^2)}}  \bigg[ 2 H(0;y) \bigg]
       + \frac{1}{2}
       + H(0,0;y)
%%%%%%%
       \bigg\} \nonumber \\
  &+& \hspace*{-0.3cm}
        \ln \Big( \frac{\lambda}{m} \Big) 
          \bigg\{ 
%%%% 
       \frac{m^2}{Q^2}   \bigg[
          - \frac{1}{12} \ \pi^2 H(0;y) 
          + 2 H(0,0;y)
          - 2 H(-1,0,0;y)
          - H(0,-1,0;y)
            \nonumber \\ 
            & & \hspace*{2.0cm}
          + \frac{3}{2} H(0,0,0;y)
          \bigg]
         \nonumber \\
  &  & 
%%%%
       \qquad \quad \ 
       + \frac{Q^2}{\sqrt{Q^2 (Q^2+4m^2)}}   \bigg[
          - \frac{1}{12} \ \pi^2
          + \frac{7}{4} H(0;y)
          - H(-1,0;y)
          + \frac{1}{2} H(0,0;y)
          \bigg]
         \nonumber \\
  &  &
%%%%
       \qquad \quad \ 
       + \frac{m^2}{Q^2+4m^2}   \bigg[
            \frac{1}{12} \ \pi^2 H(0;y) 
          - H(0,0;y)
          + 2 H(-1,0,0;y)
            \nonumber \\ 
            & & \hspace*{3.2cm}
          + H(0,-1,0;y) 
          - \frac{3}{2} H(0,0,0;y)
          \bigg]
         \nonumber \\
  &  &
%%%%
       \qquad \quad \ 
       + \frac{m^2}{\sqrt{Q^2 (Q^2+4m^2)}}   \bigg[
          - \frac{1}{6} \ \pi^2
          + 4 H(0;y)
          - 2 H(-1,0;y)
          + H(0,0;y)
          \bigg]
         \nonumber \\
  &  &
%%%%
       \qquad \quad \ 
          + 1
          - \frac{1}{12} \ \pi^2 H(0;y) 
          + \frac{3}{2} H(0,0;y)
          - 2 H(-1,0,0;y)
          - H(0,-1,0;y)
            \nonumber \\ 
            & & \qquad \quad \ 
          + \frac{3}{2} H(0,0,0;y)
%%%%%%%%%%
         \bigg\} \nonumber \\
  &+& \hspace*{-0.3cm}
        \frac{m^2}{Q^2}  \bigg[
            \frac{61}{1440} \pi^4
          - \zeta(3)
          - \frac{1}{4} \ \pi^2 H(0;y) 
          + \frac{1}{2} \ \zeta(3) H(0;y) 
          + \zeta(3) H(1;y) 
       \nonumber \\
       &  & \hspace*{0.5cm}
          + \frac{1}{12} \ \pi^2 H(-1,0;y) 
          + 2 H(0,0;y)
          + \frac{7}{24} \pi^2 H(0,0;y) 
          + \frac{1}{6} \pi^2 H(1,0;y) 
       \nonumber \\
       &  & \hspace*{0.5cm}
          - 2 H(-1,0,0;y)
          + H(0,-1,0;y)
          - \frac{1}{2} H(0,0,0;y)
          - 2 H(0,1,0;y)
       \nonumber \\
       &  & \hspace*{0.5cm}
          + 2 H(-1,-1,0,0;y)
          + H(-1,0,-1,0;y)
          - \frac{3}{2} H(-1,0,0,0;y)
       \nonumber \\
       &  & \hspace*{0.5cm}
          - H(0,-1,0,0;y)
          - \frac{3}{2} H(0,0,-1,0;y)
          + \frac{13}{4} H(0,0,0,0;y)
       \nonumber \\
       &  & \hspace*{0.5cm}
          + H(0,0,1,0;y)
          - 2 H(1,0,-1,0;y)
          + 2 H(1,0,0,0;y)
          + 2 H(1,0,1,0;y)
          \bigg]
         \nonumber \\
  &+& \hspace*{-0.3cm}
        \frac{Q^2}{\sqrt{Q^2 (Q^2+4m^2)}}  \bigg[
          - \frac{1}{32} \pi^2
          - \frac{1}{12} \pi^4
          + \frac{1627}{864} H(0;y)
          + \frac{29}{72} \pi^2 H(0;y) 
       \nonumber \\
       &  & \hspace*{2.5cm}
          - 2 \zeta(3) H(0;y)
          + \frac{61}{8} H(-1,0;y)
          + \frac{1}{2} \pi^2 H(0,-1;y) 
          - \frac{93}{16} H(0,0;y)
       \nonumber \\
       &  & \hspace*{2.5cm}
          - \frac{1}{3} \pi^2 H(0,0;y) 
          - 4 H(1,0;y)
          + \frac{29}{12} H(0,0,0;y)
          + H(0,-1,0,0;y)
       \nonumber \\
       &  & \hspace*{2.5cm}
          + 3 H(0,0,-1,0;y)
          - \frac{5}{2} H(0,0,0,0;y)
          - 2 H(0,0,1,0;y)
       \nonumber \\
       &  & \hspace*{2.5cm}
          + H(0,1,0,0;y)
          \bigg]
         \nonumber \\
  &+&  \hspace*{-0.3cm}
        \frac{m^6}{(Q^2+4m^2)^2 \ \sqrt{Q^2 (Q^2+4m^2)}}   \bigg[
            \frac{19}{8} \pi^4
          + 7 \pi^2 H(0;y) 
          - 84 \zeta(3) H(0;y) 
       \nonumber \\
       &  & \hspace*{4.7cm}
          - 15 \pi^2 H(0,-1;y) 
          + \frac{11}{2} \pi^2 H(0,0;y) 
       \nonumber \\
       &  & \hspace*{4.7cm}
          + 12 \pi^2 H(1,0;y) 
          + 42 H(0,0,0;y)
       \nonumber \\
       &  & \hspace*{4.7cm}
          + 126 H(0,-1,0,0;y)
          - 192 H(0,0,-1,0;y)
       \nonumber \\
       &  & \hspace*{4.7cm}
          + 3 H(0,0,0,0;y)
          + 48 H(0,0,1,0;y)
       \nonumber \\
       &  & \hspace*{4.7cm}
          - 180 H(0,1,0,0;y)
          + 72 H(1,0,0,0;y)
          \bigg]
         \nonumber \\
  &+&  \hspace*{-0.3cm}
        \frac{m^4}{(Q^2+4m^2)^2}   \bigg[
          - \frac{41}{9} \pi^2
          + 42 \zeta(3)
          + \frac{15}{2} \pi^2 H(-1;y) 
          + \frac{1}{4} \pi^2 H(0;y) 
          + \frac{494}{9} H(0,0;y)
       \nonumber \\
       &  & \hspace*{2.0cm}
          - 63 H(-1,0,0;y)
          + 96 H(0,-1,0;y)
          + \frac{33}{2} H(0,0,0;y)
       \nonumber \\
       &  & \hspace*{2.0cm}
          - 24 H(0,1,0;y)
          + 90 H(1,0,0;y)
          \bigg]
         \nonumber \\
  &+& \hspace*{-0.3cm}
        \frac{m^4}{(Q^2+4m^2) \ \sqrt{Q^2 (Q^2+4m^2)}} \bigg[
            4 \pi^2
          - \frac{49}{80} \pi^4
          + \frac{178}{9} H(0;y)
          - \frac{7}{4} \pi^2 H(0;y) 
       \nonumber \\
       &  & \hspace*{4.4cm}
          + 35 \zeta(3) H(0;y) 
          + 96 H(-1,0;y)
          + 4 \pi^2 H(0,-1;y) 
       \nonumber \\
       &  & \hspace*{4.4cm}
          - 60 H(0,0;y)
          - \frac{25}{12} \pi^2 H(0,0;y) 
          - 24 H(1,0;y)
       \nonumber \\
       &  & \hspace*{4.4cm}
          - 5 \pi^2 H(1,0;y) 
          - \frac{21}{2} H(0,0,0;y)
       \nonumber \\
       &  & \hspace*{4.4cm}
          - 38 H(0,-1,0,0;y)
          + 37 H(0,0,-1,0;y)
       \nonumber \\
       &  & \hspace*{4.4cm}
          + H(0,0,0,0;y)
          - 4 H(0,0,1,0;y)
       \nonumber \\
       &  & \hspace*{4.4cm}
          + 35 H(0,1,0,0;y)
          - 30 H(1,0,0,0;y)
          \bigg]
         \nonumber \\
  &+& \hspace*{-0.3cm} 
        \frac{m^2}{Q^2+4m^2}   \bigg[
          - \frac{49}{9}
          + 3 \pi^2 \ln2
          - \frac{17}{12} \pi^2
          - \frac{61}{1440} \pi^4
          - \frac{11}{2} \zeta(3)
          - \frac{15}{4} \pi^2 H(-1;y) 
       \nonumber \\
       &  & \hspace*{1.5cm}
          + \frac{35}{24} \pi^2 H(0;y)
          - \frac{1}{2} \zeta(3) H(0;y) 
          - \zeta(3) H(1;y) 
          - \frac{1}{12} \pi^2 H(-1,0;y) 
       \nonumber \\
       &  & \hspace*{1.5cm}
          - \frac{71}{3} H(0,0;y)
          - \frac{7}{24} \pi^2 H(0,0;y) 
          - \frac{1}{6} \pi^2 H(1,0;y) 
          + \frac{7}{2} H(-1,0,0;y)
       \nonumber \\
       &  & \hspace*{1.5cm}
          - 15 H(0,-1,0;y)
          + 4 H(0,0,0;y)
          + 5 H(0,1,0;y)
          - \frac{17}{2} H(1,0,0;y)
       \nonumber \\
       &  & \hspace*{1.5cm}
          - 2 H(-1,-1,0,0;y)
          - H(-1,0,-1,0;y)
          + \frac{3}{2} H(-1,0,0,0;y)
       \nonumber \\
       &  & \hspace*{1.5cm}
          + H(0,-1,0,0;y)
          + \frac{3}{2} H(0,0,-1,0;y)
          - \frac{13}{4} H(0,0,0,0;y)
       \nonumber \\
       &  & \hspace*{1.5cm}
          - H(0,0,1,0;y)
          + 2 H(1,0,-1,0;y)
          - 2 H(1,0,0,0;y)
       \nonumber \\
       &  & \hspace*{1.5cm}
          - 2 H(1,0,1,0;y)
          \bigg]
         \nonumber \\
  &+& \hspace*{-0.3cm}
        \frac{m^2}{\sqrt{Q^2 (Q^2+4m^2)}}   \bigg[
          - \frac{2}{3} \pi^2
          - \frac{11}{90} \pi^4
          + \frac{19}{108} H(0;y)
          - \frac{23}{72} \pi^2 H(0;y) 
          - 7 H(0;y) \zeta(3)
       \nonumber \\
       &  & \hspace*{2.6cm}
          - 4 H(-1,0;y)
          + \frac{1}{2} \pi^2 H(0,-1;y) 
          + H(0,0;y)
          - \frac{1}{2} \pi^2 H(0,0;y) 
       \nonumber \\
       &  & \hspace*{2.6cm}
          - 2 H(1,0;y)
          + \frac{1}{3} \pi^2 H(1,0;y) 
          - \frac{23}{12} H(0,0,0;y)
          + 5 H(0,-1,0,0;y)
       \nonumber \\
       &  & \hspace*{2.6cm}
          + 3 H(0,0,-1,0;y)
          - \frac{11}{2} H(0,0,0,0;y)
          - 4 H(0,0,1,0;y)
       \nonumber \\
       &  & \hspace*{2.6cm}
          - H(0,1,0,0;y)
          + 2 H(1,0,0,0;y)
          \bigg]
         \nonumber \\
  &+& \hspace*{-0.3cm}
            \frac{1171}{216}
          - \frac{1}{2} \pi^2 \ln2 
          + \frac{7}{16} \pi^2
          + \frac{61}{1440} \pi^4
          - \frac{9}{4} \zeta(3)
          + \frac{1}{2} \pi^2 H(-1;y) 
          - \frac{5}{16} \pi^2 H(0;y) 
       \nonumber \\
  &+& \hspace*{-0.3cm}
            \frac{1}{2} \zeta(3) H(0;y) 
          + \zeta(3) H(1;y) 
          + \frac{1}{12} \pi^2 H(-1,0;y) 
          + \frac{533}{72} H(0,0;y)
          + \frac{7}{24} \pi^2 H(0,0;y) 
       \nonumber \\
  &+& \hspace*{-0.3cm}
            \frac{1}{6} \pi^2 H(1,0;y) 
          - \frac{1}{2} H(-1,0,0;y)
          + \frac{17}{4} H(0,-1,0;y)
          - \frac{9}{8} H(0,0,0;y)
          - \frac{5}{2} H(0,1,0;y)
       \nonumber \\
  &+& \hspace*{-0.3cm}
            4 H(1,0,0;y)
          + 2 H(-1,-1,0,0;y)
          + H(-1,0,-1,0;y)
          - \frac{3}{2} H(-1,0,0,0;y)
       \nonumber \\
  &-& \hspace*{-0.3cm}
            H(0,-1,0,0;y)
          - \frac{3}{2} H(0,0,-1,0;y)
          + \frac{13}{4} H(0,0,0,0;y)
          + H(0,0,1,0;y)
       \nonumber \\
  &-& \hspace*{-0.3cm}
            2 H(1,0,-1,0;y)
          + 2 H(1,0,0,0;y)
          + 2 H(1,0,1,0;y)
         \ ;
\labbel{F1} 
\end{eqnarray}

%%%%%%%%%%%%%%%%%%%%%%%%%%%%%%%%%%%%%%%%%%%%%%%%%%%%%%%%%%%%%%%%%%%%%%%%%%%%
\begin{eqnarray}
\F_2(-Q^2) & = & 
       + \ln \Big( \frac{\lambda}{m} \Big) \bigg\{ 
            \frac{m^2}{Q^2}   \bigg[ H(0,0;y) \bigg]
          + \frac{m^2}{Q^2+4m^2} \bigg[ H(0,0;y) \bigg]
          + \frac{m^2}{\sqrt{Q^2 (Q^2+4m^2)}}  \bigg[ H(0;y) \bigg]
       \bigg\}
         \nonumber \\
  &+& \hspace*{-0.3cm}
         \frac{m^4}{Q^2 \ \sqrt{Q^2 (Q^2+4m^2)}}  \bigg[
          - \frac{23}{480} \pi^4
          + \frac{3}{4} \pi^2 H(0,-1;y) 
          - \frac{7}{24} \pi^2 H(0,0;y) 
       \nonumber \\
       &  & \hspace*{3.2cm}
          + \frac{1}{2} H(0,-1,0,0;y)
          + H(0,0,-1,0;y)
          - \frac{3}{4} H(0,0,0,0;y)
          \bigg]
         \nonumber \\
  &+& \hspace*{-0.3cm}
        \frac{m^2}{Q^2}   \bigg[
          - \frac{5}{2} \zeta(3)
          + \frac{3}{8} \pi^2 H(-1;y) 
          - \frac{17}{48} \pi^2 H(0;y) 
          - \frac{3}{2} H(0,0;y)
          + \frac{5}{4} H(-1,0,0;y)
       \nonumber \\
       &  & \hspace*{0.5cm}
          - \frac{13}{8} H(0,0,0;y)
          - H(0,1,0;y)
          - H(1,0,0;y)
          \bigg]
         \nonumber \\
  &+& \hspace*{-0.3cm}
        \frac{m^6}{(Q^2+4m^2)^2 \ \sqrt{Q^2 (Q^2+4m^2)}}   \bigg[
          - \frac{19}{8} \pi^4
          - 7 \pi^2 H(0;y) 
          + 84 \zeta(3) H(0;y) 
       \nonumber \\
       &  & \hspace*{4.7cm}
          + 15 \pi^2 H(0,-1;y) 
          - \frac{11}{2} \pi^2 H(0,0;y) 
       \nonumber \\
       &  & \hspace*{4.7cm}
          - 12 \pi^2 H(1,0;y) 
          - 42 H(0,0,0;y)
       \nonumber \\
       &  & \hspace*{4.7cm}
          - 126 H(0,-1,0,0;y)
          + 192 H(0,0,-1,0;y)
       \nonumber \\
       &  & \hspace*{4.7cm}
          - 3 H(0,0,0,0;y)
          - 48 H(0,0,1,0;y)
       \nonumber \\
       &  & \hspace*{4.7cm}
          + 180 H(0,1,0,0;y)
          - 72 H(1,0,0,0;y)
          \bigg]
         \nonumber \\
  &+& \hspace*{-0.3cm}
         \frac{m^4}{(Q^2+4m^2)^2}   \bigg[
            \frac{13}{3} \pi^2
          - 42 \zeta(3)
          - \frac{15}{2} \pi^2 H(-1;y) 
          - \frac{1}{4} \pi^2 H(0;y) 
          - \frac{166}{3} H(0,0;y)
       \nonumber \\
       &  & \hspace*{2.0cm}
          + 63 H(-1,0,0;y)
          - 96 H(0,-1,0;y)
          - \frac{33}{2} H(0,0,0;y)
       \nonumber \\
       &  & \hspace*{2.0cm}
          + 24 H(0,1,0;y)
          - 90 H(1,0,0;y)
          \bigg]
         \nonumber \\
  &+& \hspace*{-0.3cm}
        \frac{m^4}{(Q^2+4m^2) \ \sqrt{Q^2 (Q^2+4m^2)}}  \bigg[
          - 4 \pi^2
          + \frac{199}{480} \pi^4
          - \frac{62}{3} H(0;y)
          + \frac{7}{6} \pi^2 H(0;y) 
       \nonumber \\
       &  & \hspace*{4.5cm}
          - 28 \zeta(3) H(0;y) 
          - 96 H(-1,0;y)
          - \frac{11}{4} \pi^2 H(0,-1;y) 
       \nonumber \\
       &  & \hspace*{4.5cm}
          + 60 H(0,0;y)
          + \frac{13}{8} \pi^2 H(0,0;y) 
          + 24 H(1,0;y)
       \nonumber \\
       &  & \hspace*{4.5cm}
          + 4 \pi^2 H(1,0;y) 
          + 7 H(0,0,0;y)
          + \frac{55}{2} H(0,-1,0,0;y)
       \nonumber \\
       &  & \hspace*{4.5cm}
          - 21 H(0,0,-1,0;y)
          - \frac{5}{4} H(0,0,0,0;y)
       \nonumber \\
       &  & \hspace*{4.5cm}
          - 20 H(0,1,0,0;y)
          + 24 H(1,0,0,0;y)
          \bigg]
         \nonumber \\
  &+& \hspace*{-0.3cm}
        \frac{m^2}{Q^2+4m^2}  \bigg[
            \frac{17}{3}
          - 2 \pi^2 \ln2 
          + \frac{1}{6} \pi^2
          + \frac{7}{2} \zeta(3)
          + \frac{17}{8} \pi^2 H(-1;y) 
          - \frac{47}{48} \pi^2 H(0;y) 
       \nonumber \\
       &  & \hspace*{1.5cm}
          + \frac{95}{6} H(0,0;y)
          - \frac{1}{4} H(-1,0,0;y)
          + 3 H(0,-1,0;y)
          - \frac{19}{8} H(0,0,0;y)
       \nonumber \\
       &  & \hspace*{1.5cm}
          - H(0,1,0;y)
          - H(1,0,0;y)
          \bigg]
         \nonumber \\
  &+& \hspace*{-0.3cm}
        \frac{m^2}{\sqrt{Q^2 (Q^2+4m^2)}}  \bigg[
            \frac{3}{8} \pi^2
          + \frac{235}{72} H(0;y)
          + \frac{11}{12} \pi^2 H(0;y) 
          + \frac{41}{2} H(-1,0;y)
       \nonumber \\
       &  & \hspace*{2.5cm}
          - \frac{57}{4} H(0,0;y)
          - 8 H(1,0;y)
          + \frac{11}{2} H(0,0,0;y)
          \bigg] \ . 
\labbel{F2} 
\end{eqnarray}

%%%%%%%%%%%%%%%%%%%%%%%%%%%%%%%%%%%%%%%%%%%%%%%%%%%%%%%%%%%%%%%%%%%%%%%%%%%%
\section{Behaviour for $Q^2 \to \infty$}
\noindent
We present the behaviour of the two Form Factors for $Q^2 \to \infty$ 
(corresponding to $y \to 0$), writing for short $L = \ln(Q^2/m^2) $. 
The analytic continuation to large time-like $t$ can 
be carried out with the replacements $Q^2 = -(t+i\epsilon),\ $ 
$L = \ln(t/m^2)-i\pi$ -- so that for instance $L^2= \ln^2(t/m^2)-\pi^2 
-2i\pi\ln(t/m^2)$. 

\begin{eqnarray} 
\F_1(-Q^2) & = & 
    \quad  \ln^2 \Big( \frac{\lambda}{m} \Big)  \bigg\{ 
            \frac{1}{2}
          - L
          + \frac{1}{2} L^2
         \nonumber \\
  &  &
       \hspace{2.0cm}
       + \left( \frac{m^2}{Q^2} \right)  \bigg(
          - 2
          + 2 L
          \bigg)
         \nonumber \\
  &  &
       \hspace{2.0cm}
       + \left( \frac{m^2}{Q^2} \right)^2   \bigg(
            5
          - 5 L
          + 2 L^2
          \bigg)
         \nonumber \\
  &  &
       \hspace{2.0cm}
       + \left( \frac{m^2}{Q^2} \right)^3   \bigg(
          - \frac{50}{3}
          + \frac{68}{3} L
          - 8 L^2
          \bigg)
         \nonumber \\
  &  &
       \hspace{2.0cm}
       + \left( \frac{m^2}{Q^2} \right)^4   \bigg(
            \frac{196}{3}
          - \frac{183}{2} L
          + 32 L^2
          \bigg)
     \bigg\}
         \nonumber \\
  &  &
  + \ln \Big( \frac{\lambda}{m} \Big)   \bigg\{ 
            1
          - \frac{1}{12} \pi^2
          - \frac{7}{4} L
          + \frac{1}{12} \pi^2 L 
          + L^2
          - \frac{1}{4} L^3
         \nonumber \\
  &  &
       \hspace{1.8cm}
       + \left( \frac{m^2}{Q^2} \right)  \bigg(
          - \frac{5}{2}
          + \frac{1}{6} \pi^2
          + \frac{7}{2} L
          - 2 L^2
          \bigg)
         \nonumber \\
  &  &
       \hspace{1.8cm}
       + \left( \frac{m^2}{Q^2} \right)^2   \bigg(
            6
          - \frac{5}{12} \pi^2
          - \frac{55}{4} L
          + \frac{1}{3} \pi^2 L 
          + \frac{29}{4} L^2
          - L^3
          \bigg)
         \nonumber \\
  &  &
       \hspace{1.8cm}
       + \left( \frac{m^2}{Q^2} \right)^3   \bigg(
          - \frac{212}{9}
          + \frac{17}{9} \pi^2
          + \frac{500}{9} L
          - \frac{4}{3} \pi^2 L 
          - \frac{97}{3} L^2
          + 4 L^3
          \bigg)
         \nonumber \\
  &  &
       \hspace{1.8cm}
       + \left( \frac{m^2}{Q^2} \right)^4   \bigg(
            \frac{13813}{144}
          - \frac{61}{8} \pi^2
          - \frac{11101}{48} L
          + \frac{16}{3} \pi^2 L 
          + \frac{1079}{8} L^2
          - 16 L^3
          \bigg)
     \bigg\}
         \nonumber \\
  &  &
  \nonumber \\
  &  &
          + \frac{1171}{216}
          - \frac{1}{2} \pi^2 \ln2 
          + \frac{13}{32} \pi^2
          - \frac{59}{1440} \pi^4
          - \frac{9}{4} \zeta(3)
          - \frac{1627}{864} L
          - \frac{13}{144} \pi^2 L 
          + \frac{3}{2} \zeta(3) L 
          \nonumber \\
          &  &
          \nonumber \\
          &  &
          + \frac{229}{288} L^2
          - \frac{1}{48} \pi^2 L^2 
          - \frac{31}{144} L^3
          + \frac{1}{32} L^4
         \nonumber \\
  &  &
  \nonumber \\
  &  &
       + \left( \frac{m^2}{Q^2} \right)  \bigg(
          - \frac{5113}{432}
          + 3 \pi^2 \ln2 
          - \frac{209}{144} \pi^2
          + \frac{2}{45} \pi^4
          - \frac{5}{2} \zeta(3)
          + \frac{283}{48} L
          - \frac{5}{12} \pi^2 L 
          \nonumber \\
          &  & 
          \hspace*{1.8cm}
          + 3 \zeta(3) L 
          - \frac{61}{16} L^2
          + \frac{1}{12} \pi^2 L^2 
          + \frac{17}{24} L^3
          - \frac{1}{48} L^4
          \bigg)
         \nonumber \\
  &  &
       + \left( \frac{m^2}{Q^2} \right)^2   \bigg(
            \frac{11899}{144}
          - 12 \pi^2 \ln2 
          - \frac{101}{144} \pi^2
          - \frac{503}{720} \pi^4
          + 64 \zeta(3)
          - \frac{3977}{216} L
          + \frac{175}{36} \pi^2 L 
          \nonumber \\
          &  & 
          \hspace*{2.0cm}
          - 39 \zeta(3) L 
          + \frac{3541}{144} L^2
          - \frac{23}{24} \pi^2 L^2 
          - \frac{43}{18} L^3
          + \frac{5}{12} L^4
          \bigg)
         \nonumber \\
  &  &
       + \left( \frac{m^2}{Q^2} \right)^3   \bigg(
          - \frac{272405}{432}
          + 48 \pi^2 \ln2 
          + \frac{9929}{216} \pi^2
          + \frac{227}{36} \pi^4
          - \frac{1481}{3} \zeta(3)
          - \frac{1655}{108} L
          \nonumber \\
          &  & 
          \hspace*{2.0cm}
          - \frac{1061}{36} \pi^2 L 
          + 304 \zeta(3) L 
          - \frac{425}{3} L^2
          + \frac{17}{2} \pi^2 L^2 
          + \frac{133}{9} L^3
          - \frac{19}{12} L^4
          \bigg)
         \nonumber \\
  &  &
       + \left( \frac{m^2}{Q^2} \right)^4   \bigg(
            \frac{82660981}{20736}
          - 192 \pi^2 \ln2 
          - \frac{689557}{1728} \pi^2
          - \frac{15409}{360} \pi^4
          + 3056 \zeta(3)
          \nonumber \\
          &  & 
          \hspace*{2.0cm}
          + \frac{446243}{864} L
          + \frac{3995}{24} \pi^2 L 
          - 1922 \zeta(3) L 
          + \frac{435125}{576} L^2
          - \frac{673}{12} \pi^2 L^2 
          \nonumber \\
          &  & 
          \hspace*{2.0cm}
          - \frac{155}{6} L^3
          + \frac{143}{24} L^4
          \bigg)
         \nonumber \\
  &  &
          + \mathcal{O} \left( \ \left( \frac{m^2}{Q^2} \right)^5 \ \right) \ .
\labbel{F1towardinf}
\end{eqnarray}
\noindent 
The above formula matches with the sum of Eq.s(2.23) and (2.34-2.35a) 
of~\cite{LEP}, which corresponds to 
the real part of $\F_1(t)$ for large timelike $t$ expanded in $m^2/t$ 
up to the zeroth order. 
%%%%%%%%%%%%%%%%%%%%%%%%%%%%%%%%%%%%%%%%%%%%%%%%%%%%%%%%%%%%%%%%%%%%%%%%%%%%
\begin{eqnarray}
\F_2(-Q^2) & = & 
     \ln \Big( \frac{\lambda}{m} \Big)   \bigg\{ 
         \left( \frac{m^2}{Q^2} \right)  \bigg(
          - L
          + L^2
          \bigg)
         \nonumber \\
  &  &
       \hspace{1.8cm}
       + \left( \frac{m^2}{Q^2} \right)^2   \bigg(
            2
          + 6 L
          - 2 L^2
          \bigg)
         \nonumber \\
  &  &
       \hspace{1.8cm}
       + \left( \frac{m^2}{Q^2} \right)^3   \bigg(
            11
          - 20 L
          + 8 L^2
          \bigg)
         \nonumber \\
  &  &
       \hspace{1.8cm}
       + \left( \frac{m^2}{Q^2} \right)^4   \bigg(
          - \frac{134}{3}
          + \frac{232}{3} L
          - 32 L^2
          \bigg)
    \bigg\}
         \nonumber \\
  &  &
       + \left( \frac{m^2}{Q^2} \right)  \bigg(
            \frac{17}{3}
          - 2 \pi^2 \ln2 
          + \frac{13}{24} \pi^2
          + \zeta(3)
          - \frac{235}{72} L
          + \frac{5}{12} \pi^2 L 
          + \frac{1}{24} L^2
          - \frac{1}{4} L^3
          \bigg)
         \nonumber \\
  &  &
       + \left( \frac{m^2}{Q^2} \right)^2   \bigg(
          - \frac{1609}{36}
          + 8 \pi^2 \ln2 
          + \frac{9}{4} \pi^2
          + \frac{11}{30} \pi^4
          - 56 \zeta(3)
          + \frac{463}{36} L
          - 3 \pi^2 L 
          \nonumber \\
          &  & 
          \hspace*{2.0cm}
          + 28 \zeta(3) L 
          - \frac{205}{12} L^2
          + \frac{2}{3} \pi^2 L^2 
          + \frac{11}{6} L^3
          - \frac{1}{12} L^4
          \bigg)
         \nonumber \\
  &  &
       + \left( \frac{m^2}{Q^2} \right)^3   \bigg(
            \frac{33121}{72}
          - 32 \pi^2 \ln2 
          - \frac{165}{4} \pi^2
          - \frac{143}{30} \pi^4
          + 448 \zeta(3)
          + \frac{19}{12} L
          \nonumber \\
          &  & 
          \hspace*{2.0cm}
          + \frac{125}{6} \pi^2 L 
          - 252 \zeta(3) L 
          + \frac{479}{4} L^2
          - \frac{22}{3} \pi^2 L^2 
          - \frac{71}{6} L^3
          + \frac{1}{4} L^4
          \bigg)
         \nonumber \\
  &  &
       + \left( \frac{m^2}{Q^2} \right)^4   \bigg(
          - \frac{87919}{27}
          + 128 \pi^2 \ln2 
          + \frac{3154}{9} \pi^2
          + \frac{359}{10} \pi^4
          - 2828 \zeta(3)
          - \frac{6589}{18} L
          \nonumber \\
          &  & 
          \hspace*{2.0cm}
          - \frac{386}{3} \pi^2 L 
          + 1680 \zeta(3) L 
          - \frac{4085}{6} L^2
          + 51 \pi^2 L^2 
          + 65 L^3
          - \frac{1}{2} L^4
          \bigg)
         \nonumber \\
  &  &
          + \mathcal{O} \left( \ \left( \frac{m^2}{Q^2} \right)^5 \ \right) \ .
\labbel{F2towardinf}
\end{eqnarray}
%%%%%%%%%%%%%%%%%%%%%%%%%%%%%%%%%%%%%%%%%%%%%%%%%%%%%%%%%%%%%%%%%%%%%%%%%%%%

\section{Behaviour for $Q^2 \to 0$}
\noindent
We present the behaviour of the two Form Factors around $Q^2=0$, 
corresponding to $y = 1$. 

\begin{eqnarray}
\F_1(-Q^2) & = & 
       \quad \ln^2 \Big( \frac{\lambda}{m} \Big) \bigg\{ 
           \bigg( \frac{Q^2}{m^2} \bigg)^2   \bigg(\frac{1}{18} \bigg)
         + \bigg( \frac{Q^2}{m^2} \bigg)^3  \bigg(-\frac{1}{60}\bigg)
         \bigg\}
         \nonumber \\
  &  &
       + \ln \Big( \frac{\lambda}{m} \Big) \bigg\{
           \bigg( \frac{Q^2}{m^2} \bigg)^2  \bigg(\frac{1}{24}\bigg)
         + \bigg( \frac{Q^2}{m^2} \bigg)^3  \bigg(-\frac{31}{1440}\bigg)
         \bigg\}
         \nonumber \\
  &  &
       + \bigg( \frac{Q^2}{m^2} \bigg)   \bigg(
            \frac{4819}{5184}
          - \frac{1}{2} \pi^2 \ln2 
          + \frac{49}{432} \pi^2
          + \frac{3}{4} \zeta(3)
          \bigg)
         \nonumber \\
  &  &
       + \bigg( \frac{Q^2}{m^2} \bigg)^2   \bigg(
          - \frac{1889}{6480}
          + \frac{11}{60} \pi^2 \ln2 
          - \frac{8731}{172800} \pi^2
          - \frac{11}{40} \zeta(3)
          \bigg)
         \nonumber \\
  &  &
       + \bigg( \frac{Q^2}{m^2} \bigg)^3   \bigg(
            \frac{163249}{1814400}
          - \frac{113}{1680} \pi^2 \ln2 
          + \frac{723901}{33868800} \pi^2
          + \frac{113}{1120} \zeta(3)
          \bigg)
         \nonumber \\
  &  &
          + \mathcal{O} \Bigg( \bigg( \frac{Q^2}{m^2} \bigg)^4 \Bigg) \ .
\labbel{F1to0}
\end{eqnarray}
%%%%%%%%%%%%%%%%%%%%%%%%%%%%%%%%%%%%%%%%%%%%%%%%%%%%%%%%%%%%%%%%%%%%%%%%%%%%

\begin{eqnarray}
\F_2(-Q^2) & = & 
       \quad \ln \Big( \frac{\lambda}{m} \Big)   \bigg\{
         \bigg( \frac{Q^2}{m^2} \bigg)   \bigg( \frac{1}{6} \bigg)
       + \bigg( \frac{Q^2}{m^2} \bigg)^2  \bigg( - \frac{19}{360} \bigg)
       + \bigg( \frac{Q^2}{m^2} \bigg)^3  \bigg(  \frac{73}{5040} \bigg)
       \bigg\}
          \nonumber \\
 &  &
       + \frac{197}{144}
       - \frac{1}{2} \pi^2 \ln2 
       + \frac{1}{12} \pi^2
       + \frac{3}{4} \zeta(3)
          \nonumber \\
 &  &
       + \bigg( \frac{Q^2}{m^2} \bigg)   \bigg(
          - \frac{1751}{2160}
          + \frac{23}{60} \pi^2 \ln2 
          - \frac{13}{120} \pi^2
          - \frac{23}{40} \zeta(3)
          \bigg)
          \nonumber \\
 &  &
       + \bigg( \frac{Q^2}{m^2} \bigg)^2   \bigg(
            \frac{4357}{15120}
          - \frac{5}{28} \pi^2 \ln2 
          + \frac{187}{3150} \pi^2
          + \frac{15}{56} \zeta(3)
          \bigg)
          \nonumber \\
 &  &
       + \bigg( \frac{Q^2}{m^2} \bigg)^3   \bigg(
          - \frac{111619}{1209600}
          + \frac{29}{420} \pi^2 \ln2 
          - \frac{140951}{5644800} \pi^2 
          - \frac{29}{280} \zeta(3)
          \bigg)
          \nonumber \\
 &  &
          + \mathcal{O} \Bigg( \bigg( \frac{Q^2}{m^2} \bigg)^4 \Bigg) \ .
\labbel{F2to0}
\end{eqnarray}
From the above expansions one can easily recover some familiar 
results, such as the renormalization condition $\F_1(0) = 0$ and 
the two loop values for the slope ${F_1^{(2)}}'(0)$ (the coefficient of 
the first term in $Q^2$ in Eq.(\ref{eq:F1to0}) with an overall minus sign) 
and of the electron anomaly $\F_2(0)$. 
%%%%%%%%%%%%%%%%%%%%%%%%%%%%%%%%%%%%%%%%%%%%%%%%%%%%%%%%%%%%%%%%%%%%%%%%%%%%
\section{Acknowledgments} 

We are grateful to J. Vermaseren for his assistance in the use of 
of FORM \cite{FORM} by which all the algebra was processed. 
We acknowledge the interest of U.D. Jentschura~\cite{Jents} for 
the analytic expressions of the derivatives of the form factors 
at zero momentum transfer, which initiated the present work, 
the clarifying remarks of R. Bonciani on the asymptotic 
expansions, and useful discussions with H. Czy\.z and  J.H. K\"uhn. 
%%%%%%%%%%%%%%%%%%%%%%%%%%%%%%%%%%%%%%%%%%%%%%%%%%%%%%%%%%%%%%%%%%%%%%%%%%%%
\appendix

\section{The Harmonic Polylogarithms, HPL's}
\setcounter{equation}{0}
\renewcommand{\theequation}{A-\arabic{equation}}

We recall for convenience of the reader the definition of the 
HPL's~\cite{hpl}. 
The HPL's form a family of functions depending on an argument, say $x$, 
and on a set of indices, say $a_i, i=1,..,w$ or $\vec a$ in more compact 
notation, where each of the $a_i$ can take one of the three values $1, 0, -1$ 
and whose number $w$ is called the weight of the HPL. At weight $w=1$ 
there are 3 HPL's, defined as 
\begin{eqnarray} 
  H(1;x) &=& - \ln(1-x) \ , \nonumber\\ 
  H(0;x) &=& \ln{x} \ , \nonumber\\ 
  H(-1;x) &=& \ln(1+x) \ , 
\label{eq:defw1} 
\end{eqnarray}
whose derivatives can be written as 
\begin{equation} 
  \frac{d}{dx}H(a;x) = f(a;x) \ , 
\label{eq:defdHw1} 
\end{equation} 
where the index $a$ can take one of the three values $(1,0,-1)$ and the 
rational factors $f(a;x)$ are given by 
\begin{eqnarray} 
  f(1;x)  &=& \frac{1}{1-x} \ , \nonumber\\ 
  f(0;x)  &=& \frac{1}{x} \ , \nonumber\\ 
  f(-1;x) &=& \frac{1}{1+x} \ . 
\label{eq:deff} 
\end{eqnarray} 
\par 
At weight $w>1$, if all the $w$ indices are equal to $0$ let us indicate 
them by $\vec 0_w$ and define correspondingly 
\begin{equation} 
   H(\vec 0_w;x) = \frac{1}{w!} \ln^w{x} \ ; 
\labbel{H0w} 
\end{equation} 
in all the other cases (\ie when the indices are not all equal to zero), 
let us indicate any set of $w$ indices by $(a,\vec b)$, where $a$ can 
take one of the three values $(1,0,-1)$ and $\vec b$ stands for the set 
of the other $w-1$ indices, and define correspondingly 
\begin{equation} 
   H(a,\vec b;x) = \int_0^x dx' f(a;x') H(\vec b;x') \ . 
\labbel{Hw} 
\end{equation} 
Note that in full generality (\ie also when all the indices are equal to 0) 
one has 
\begin{equation} 
  \frac{d}{dx}H(a,\vec b;x) = f(a;x) H(\vec b;x) \ , 
\labbel{dxHw} 
\end{equation} 
which can also be written as the equivalent indefinite integration formula 
\begin{equation} 
   \int^x dx' f(a;x') \ H(\vec b;x') = A + \ H(a,\vec b;x) \ , 
\labbel{intdxHw} 
\end{equation} 
where $A$ is an integration constant. \par 
Further, the product of two HPL's of the a same argument $x$ 
and weights $p, q$ can be expressed as a combination of HPL's  of 
argument $x$ and weight $r=p+q$, according to the product identity 
\begin{eqnarray} 
 H(\vec{p};x) H(\vec{q};x) & = & 
  \sum_{\vec{r} = \vec{p}\uplus \vec{q}} \ H(\vec{r};x) \; , 
\labbel{algebra} \end{eqnarray} 
where $\vec p, \vec q$ stand for the $p$ and $q$ components of the indices 
of the two HPL's, while $\vec{p}\uplus \vec{q}$ represents all possible 
mergers of $\vec{p}$ and $\vec{q}$ into the vector $\vec{r}$ with $r$ 
components, in which the relative orders of the elements of $\vec{p}$ 
and $\vec{q}$ are preserved. The simplest cases of the above identities 
are 
\begin{eqnarray} 
 \ H(a;x) \ H(b;x) &=& \ H(a,b;x) + \ H(b,a;x) \ , \nonumber\\ 
 \ H(a;x) \ H(b,c;x) &=& \ H(a,b,c;x) + \ H(b,a,c;x) + \ H(b,c,a;x) \ ; 
\labbel{algebraex} \end{eqnarray} 
more complicated cases are immediately established recursively (all the above 
formulae can indeed easily be checked by differentiating, repeatedly when 
needed, with respect to $x$). 
All the $3^w$ HPL's of weight $w$ are linearly independent; 
Eq.(\ref{eq:algebraex}) can however be used for replacing an HPL of weight $w$ 
with products of HPL's of lower weight (but such that the sum of their 
weights is equal to $w$) and other HPL's of weight $w$. \par 
In the analytic continuation from the argument $y$ in the range $(0,1)$ 
to $y=-x+i\epsilon$, where $x$ is again in the range $(0,1)$, one 
finds that any HPL with rightmost index equal to $0$ does develop an 
imaginary part. It is therefore convenient to exploit Eq.(\ref{eq:algebra}) 
for expressing those HPL's in terms of HPL's having either no $0$'s on 
the right or only $0$'s as indices. One has for instance 
\begin{equation} 
  H(-1,0;y) = H(-1;y) H(0;y) - H(0,-1;y) \ , \nonumber 
\end{equation} 
and similar formulae for more general cases. From the very definition 
one has further $ H(0;-x+i\epsilon)= H(0;x)+i\pi,\ $ and 
$ H(-1;-x+i\epsilon) = H(-1;-x) = -H(1;x),\ $ 
$ H(0,-1;-x+i\epsilon) = - H(0,1;x)\ ,$ (those last two functions have 
no imaginary part in that range of values of $x$), so that finally 
one obtains 
\begin{equation} 
 H(-1,0;-x+i\epsilon) = - H(0;x)H(1;x) + H(0,1;x) + i\pi H(1;x) \ . 
\nonumber \end{equation} 
For a more complete discussion see~\cite{hpl,num1hpl}. 
\par 
Of particular interest are the values of the HPL's of argument equal to 1. 
They can be expressed as combination with rational factors of a limited 
number of mathematical constants, such as for instance Riemann 
$\zeta$-functions, $\pi^2$ and $\ln2$. 
%%%%%%%%%%%%%%%%%%%%%%%%%%%%%%%%%%%%%%%%%%%%%%%%%%%%%%%%%%%%%%%%%%%%%%%%%%%%%
\section{The definite integrals occurring in the calculation}
\setcounter{equation}{0}
\renewcommand{\theequation}{B-\arabic{equation}}

The imaginary parts of the form factors at two loops 
contain Nielsen's polylogarithms of maximum weight 3 and arguments 
$x', -x', x'^2$, which can be expressed in terms of HPL's of argument $x'$. 
Indeed, one has $Li_2(x)=H(0,1;x)$, $\ Li_2(-x)=-H(0,-1;x)$, 
$\ Li_3(x)=H(0,0,1;x)$, $\ Li_3(-x)=-H(0,0,-1;x)$, 
$\ S_{12}(x)=H(0,1,1;x)$, $\ S_{12}(-x)=H(0,-1,-1;x)$ and 
$$ S_{12}(x^2) = 2H(0,1,1;x)-2H(0,1,-1;x)-2H(0,-1,1;x)+2H(0,-1,-1;x). $$ 
After the insertion of the explicit analytic expressions of the imaginary 
parts in Eq.s(\ref{eq:ReF1},\ref{eq:ReF2}) and full partial fractioning 
in the integration variable $x'$, the integrands are found to consist of 
powers of the rational factors $x',\ 1/(1+x'),\ 1/(1-x'),\ $ 
which depend on $x'$ only, and $1/(x'+y),\ $ $1/(x'+1/y),\ $ 
which depend as well on $y$, times HPL's of argument $x'$ and maximum 
weight $w=3$. 
\par 
The integrals with the factors depending on $x'$ only can be evaluated 
by parts or by means of the formulas defining the HPL's; they give 
end-point values of HPL's of argument 1 and maximum weight $w=4$ (some 
care may be needed in properly grouping terms whose end-point values are 
otherwise separately divergent). \par 
For the analytic evaluation of the integrals involving $1/(x'+y)$ and 
$1/(x'+1/y)$, along the lines of Section 7 of~\cite{hpl}, we introduce 
two families of related functions depending, as the HPL's, on a 
set of $w$ indices and the argument $y$, defined for $w=1$ as 
\begin{eqnarray} 
 F(-1;y) &:=& \int_0^1 \frac{dx'}{x'+y} = H(-1;y) - H(0;y) \ , 
                                                      \labbel{defF1} \\ 
 G(-1;y) &:=& \int_0^1 \frac{dx'}{x'+\frac{1}{y}} = H(-1;y) \ , 
                                                      \labbel{defG1} 
\end{eqnarray} 
and for $w>1$ as 
\begin{eqnarray} 
 F(-1,\vec b;y) &:=& \int_0^1 \frac{dx'}{x'+y} H(\vec b;x') \ , 
                                                         \labbel{defFw} \\ 
 G(-1,\vec b;y) &:=& \int_0^1 \frac{dx'}{x'+\frac{1}{y}} H(\vec b;x') \ . 
                                                         \labbel{defGw} 
\end{eqnarray} 
Note that the first index of the above functions is frozen to the value 
$-1$ (a convention suggested by the $y=1$ limiting value). \par 
It is easy to see, proceeding by induction on $w$, that the functions 
$F(-1,\vec a;y)$ and $G(-1,\vec a ;y)$ with a total of $w$ indices are 
just homogeneous combination of HPL's functions of weight $w$ of argument 
$y$ and of their values at $1$. \par 
For $w = 1$, the result is obvious from 
the definitions Eq.s(\ref{eq:defF1},\ref{eq:defG1}). 
Next, assume that the identities are established for $F(-1,\vec{b};y)$ 
up to a certain weight $w-1\ge1$ and consider the function of weight $w$ 
\begin{equation} 
  F(-1,a,\vec b;y) = \int_0^1 \frac{dx'}{x'+y} H(a,\vec b;x') \ . 
\labbel{Fwp1} 
\end{equation} 
Its values at $y=0,1$, according to the definition of the HPL's 
Eq.(\ref{eq:Hw}) are 
\begin{eqnarray} 
F(-1,a,\vec b;1) &=& H(-1,a,\vec b;1) \ , \nonumber\\ 
F(-1,a,\vec b;0) &=& H(0,a,\vec b;1) \ . \labbel{Fat01} 
\end{eqnarray} 
By differentiating Eq.(\ref{eq:Fwp1}) with respect to $y$ and integrating 
by parts on $x'$ one obtains 
\begin{eqnarray}
\frac{\partial}{\partial y} F(-1,a,\vec b;y) 
    & = &   f(-1;y) H(a,\vec b; 1) 
          - \int_0^1 \frac{dx'}{x' + y}
                   \ f(a;x') \ H(\vec b;x') \ . 
\end{eqnarray}
(We have assumed $H(a,\vec b;0)=0$, which is the most frequent case; 
for $H(a,\vec b;0)\ne 0$ see the remarks below.)
\noindent
In the case $a = 0$, after partial fractioning in $x'$, one has 
\begin{eqnarray}
\frac{\partial}{\partial y} F(-1,0,\vec b;y) 
    & = &   f(-1;y) H(0,\vec b; 1) 
          - f(0;y) \int_0^1 dx' 
                    \left(f(0;x') - \frac{1}{x'+y} \right)
                     \ H(\vec b;x') 
          \nonumber \\
    & = &  f(-1;y) H(0,\vec b; 1)
          - f(0;y) \bigg( H(0,\vec b;1) - F(-1,\vec b;y) \bigg) \ ,
\end{eqnarray}
where Eq.(\ref{eq:Hw}) and Eq.(\ref{eq:defFw}) have been used to carry
out the $x'$-integration. 

For $a = 1,-1$ one obtains similarly 
\begin{eqnarray}
\frac{\partial}{\partial y} F(-1,1,\vec b;y) 
    & = & - f(-1;y) F(-1,\vec b;y) \ , \\ 
\frac{\partial}{\partial y} F(-1,-1,\vec b;y) 
    & = & f(-1;y) H(-1,\vec b; 1) 
          + f(1;y) \bigg(  H(-1,\vec{b}\,;1) - F(-1,\vec b;y) \bigg)
\end{eqnarray} 
\noindent
In the {\it r.h.s.} of the previous equations appears the function
$F(-1,\vec b;y)$ of weight $w$; when proceeding by induction in $w$ 
one can substitute the already obtained identities expressing 
it in terms of $H$'s of weight $w-1$ and argument $y$. 
A final quadrature in $y$, carried out according to Eq.(\ref{eq:Hw}), 
gives $ F(-1,a,\vec{b};y) $ in terms of HPL's of weight $w$ up to 
an additive constant, which can be fixed by one of the Eq.s(\ref{eq:Fat01}). 

It is not difficult, following the above lines, to work out all  
the formulas needed for the calculation. 
As an example, we give one of the identities of weight $w=4$, 
\begin{eqnarray}
F(-1,1,0,0;x) 
    & = &
     + \frac{1}{6} H(-1;x) H(0;x) H(0;x) H(0;x)
     + 2 H(-1;x) H(0,-1;1) H(0;x) 
            \nonumber \\
    &   & 
     - H(-1;x) H(0,0,-1;x)
     + H(-1;1) H(0,0,-1;1)
     + H(-1;1) H(0,0,1;1)
            \nonumber \\
    &   & 
     - \frac{1}{2} H(0,-1;x) H(0;x) H(0;x)
     + \frac{1}{2} H(0,-1;x) H(0,-1;x)
            \nonumber \\
    &   & 
     - 2 H(0,-1;1) H(0,-1;x)
     + \frac{3}{2} H(0,-1;1) H(0,-1;1)
     + H(0,0,-1;x) H(0;x)  \qquad 
            \nonumber \\
    &   & 
     - H(0,0,0,-1;x)
     + H(0,0,0,-1;1)
     - H(0,0,1,-1;1) \ .
\end{eqnarray}

Essentially the same procedure applies to the other family of functions 
$G(-1,\vec a ;x)$ as well. \par 
For particular values of the arguments, the end-point values 
occurring in the derivation may give divergent contributions when 
taken separately; in those cases, one can parameterize the end-point 
singularities by integrating on $x'$ from $\delta$ to $1-\eta$ 
and then take the $\delta,\eta \to 0$ limit in the final result 
(which is of course finite). 
%%%%%%%%%%%%%%%%%%%%%%%%%%%%%%%%%%%%%%%%%%%%%%%%%%%%%%%%%%%%%%%%% 

%%%%%%%%%%%%%%%%%%%%%%%%%%%%%%%%%%%%%%%%%%%%%%%%%%%%%%%%%%%%%%%%% 

\end{document}